\documentclass[12pt]{article}

\RequirePackage[OT1]{fontenc}
\RequirePackage{amsthm,amsmath,amssymb}
\RequirePackage{natbib}
\RequirePackage[colorlinks,citecolor=blue,urlcolor=blue]{hyperref}

\usepackage[all]{hypcap}
\usepackage{graphicx}
\usepackage{multirow}
\usepackage{amsmath}
\usepackage{caption}
\usepackage{amsfonts}
\usepackage{bbding}
\usepackage{float}
\usepackage{amssymb, latexsym}
\usepackage{amsthm}
\usepackage{amsthm,color}
\usepackage{pifont}% http://ctan.org/pkg/pifont

\bibpunct{(}{)}{,}{autheryear}{,}{;}

\newcommand{\blind}{1}

\newtheorem{theorem}{Theorem}

\def\no{\noindent}

\newcommand{\bx}{\mathbf{x}}
\newcommand{\bX}{\mathbf{X}}

\newcommand{\bc}{\mathbf{c}}
\newcommand{\bh}{\mathbf{h}}
\newcommand{\bm}{\mathbf{m}}

\newcommand{\bdelta}{{\mbox{\boldmath$\delta$}}}

\newcommand{\bbeta}{{\mbox{\boldmath$\beta$}}}
\newcommand{\bphi}{{\mbox{\boldmath$\phi$}}}

\newcommand{\xmark}{\ding{55}}%

\def\IPW{\scriptscriptstyle IPW}

\def\red{\color{black}}

\begin{document}

\def\spacingset#1{\renewcommand{\baselinestretch}%
{#1}\small\normalsize} \spacingset{1}

\if1\blind
{
  \title{\bf Combining Non-probability and Probability Survey Samples Through Mass Imputation}
  \author{Jae Kwang Kim$^1$,
  	 Seho Park$^2$, Yilin Chen$^3$, and Changbao Wu$^3$\\
    $^{1}$ Department of Statistics, Iowa State University,\\
     $^2$ Department of Biostatistics, Indiana University School of Medicine, \\
    $^3$ Department of Statistics and Actuarial Science,
University of Waterloo}
\date{}
  \maketitle
} \fi

\if0\blind
{
  \bigskip
  \bigskip
  \bigskip
  \begin{center}
    {\LARGE\bf Combining Non-probability and Probability Survey Samples Through Mass Imputation}
\end{center}
  \medskip
} \fi

\bigskip

\noindent%
\textbf{Summary.} Analysis of non-probability survey samples requires auxiliary information {\red at the population level. Such information may also be obtained from an existing probability survey sample from the same finite population}. 
Mass imputation has been used in practice for combining non-probability and probability survey samples {\red and making inferences on the parameters of interest  using the information collected only in the non-probability sample for the study variables.}   
%The method was originally proposed by \cite{rivers2007} as sample matching but has never been treated with rigorous theoretical justification. 
{\red Under the assumption that the conditional mean function from the non-probability sample can be transported to the probability sample}, we establish the consistency of the mass imputation estimator  and derive its asymptotic variance formula. Variance estimators are developed using either linearization or bootstrap. Finite sample performances of the mass imputation estimator are investigated through simulation studies. {\red We also address important practical issues of the method through the analysis of a real world non-probability survey sample collected by the Pew Research Centre.}

\noindent%

\vfill

\newpage
\spacingset{1} % DON'T change the spacing!

\section{Introduction}

Probability sampling is a classical tool for obtaining a representative sample from a target population. Because the first-order inclusion probabilities are known under the given survey design,  probability sampling can provide design unbiased estimators and valid statistical inferences for finite population parameters. Although probability samples have been a tremendously successful story over the past 60 years, the approach has encountered many challenges in recent years, including high cost, low response rate, and inability to provide up-to-date information on variables of specific studies.
On the other hand, non-probability samples, such as those collected from web panels, have become increasingly popular due to the efficient recruitment process, quick responses and low maintenance expenses. See, for instance, \cite{Tourangeau2013}, for examples of web based non-probability survey samples. 

Statistical analysis of non-probability survey samples, however, faces many challenges as documented by \cite{baker13}. {\red The first major challenge is that the selection/inclusion mechanism for non-probability samples is typically unknown. Treating non-probability samples as if they are a simple random sample often leads to biased results. The second major challenge is that the estimation of the propensity scores (i.e., the probabilities of participating in the non-probability sample) under an assumed model requires auxiliary information at the population level. A popular framework used in recent years is to assume that the required auxiliary information on the target population is available from an existing probability survey sample.} This framework was first used by \cite{rivers2007} and followed by a number of other authors, including \cite{Vavreck2008}, \cite{Lee2009}, \cite{Valliant2011}, \cite{Brick2015}, \cite{elliott2017inference} and \cite{chen2018doubly}, among others. 
%Combining the up-to-date information from a non-probability sample and auxiliary information from a probability sample can be viewed as data integration, which is an emerging area of research in survey sampling \citep{lohr2017combining}.

{\red Generally speaking, there are three possible approaches to analyzing non-probability survey samples: (i) inverse probability weighting (IPW) using estimated propensity scores; (ii) model-based prediction using an outcome regression model; and (iii) doubly robust procedures involving both the propensity scores and the outcome regression model. A central point to all approaches is the assumption that 
the selection/inclusion mechanism for non-probability survey samples is ignorable. It is essentially the missing at random (MAR) assumption of \cite{rubin1976}. It is well known in the literature on missing data analysis that the MAR assumption cannot be formally tested using data from the sample itself. Practical applications of the propensity score based methods require a careful examination of the auxiliary variables included in the sample to see whether participation in the non-probability sample could be characterized by those variables. Calibration weighting methods for non-probability samples are also discussed in the literature. See, for instance, \cite{dever2016} and  \cite{elliott2017inference}. The proposed methodologies, however, are closely related to the propensity scores based methods and require the same ignorability assumption on the selection mechanism \citep{chen2018doubly}. 

Mass imputation, on the other hand,  is a model-based prediction method that includes both the non-probability and the probability samples. 
%The validity of the method relies on the transportability condition which is similar to but slightly weaker than the ignorability assumption.  
The existing probability sample is assumed to have no measurement on the study variable of interest and can be viewed as having 100\% missing values for the study variable.
The non-probability sample contains  values observed on both the study variable and the auxiliary variables. 
The non-probability sample is then used as  training data to develop a prediction model that will be used to provide imputations  for the probability sample.
The key assumption for the validity of the mass imputation method is that the prediction model built based on the non-probability sample can be transported to the probability sample for imputation. See Section \ref{sec2} for detailed discussions. Mass imputation has also been developed in the context of two-phase sampling (\citealp{Breidt1996, kim2012, park2019}), but it is not fully investigated in the context of survey integration for combining the non-probability sample with a probability sample. One notable exception is the sample matching method of \cite{rivers2007}, but he did not provide any theoretical justifications for the proposed method. Furthermore, the sample matching method of  \cite{rivers2007} is based on the nearest neighbor imputation method, which suffers from the curse of dimensionality when the number of auxiliary variables is large.
\cite{chipperfield2012combining}
discussed composite estimation when one of the surveys is 
mass-imputed.
\cite{bethlehem2016solving} discussed practical issues in sample
matching for mass imputation.

In this paper, we aim to fill the important research gap in survey sampling on mass imputation for analyzing non-probability survey samples. From a theoretical point of view, if the training data for building the imputation model were a probability sample, then the theory of \cite{kim2012} could be  directly  applicable. Under the assumptions that the support of auxiliary variables from the non-probability sample is the same as the sample space and the conditional mean function from the non-probability sample can be transported to the probability sample, we show that the method of \cite{kim2012} can be applied even when  non-probability samples are used as the training data. We develop rigorous asymptotic theory for the mass imputation estimator along with a linearization variance estimator and a proposed bootstrap method for variance estimation. Our investigation on practical aspects of the method was motivated by the analysis of a real non-probability survey sample collected in 2015 by the Pew Research Centre (PRC) from the United States of America. It turned out that there existed at least two probability survey samples taken in the same year from the same finite population with some common auxiliary variables included in both samples. In addition to demonstrations of the mass imputation method using the PRC non-probability sample, we also address major practical questions through the example, which includes (i) the assessment of auxiliary variables with regard to the ignorability assumption and the imputation model; (ii) the impact of relative sample sizes between the non-probability and the probability samples; and (iii) factors to be considered in choosing a probability sample when multiple samples are available. 
}

The paper is organized as follows. 
The basic setting is described in Section \ref{sec2} and the mass imputation is presented in Section \ref{sec3}. Main theoretic results on consistency and the asymptotic variance formula are presented in Section \ref{sec4}. A practically useful bootstrap variance estimator is proposed in Section \ref{sec5}. Results from a limited simulation study on the finite sample performances of the mass imputation estimator are reported in Section \ref{sec6}.
Important practical aspects of the method are discussed in Section \ref{sec7} through an application to the PRC non-probability survey sample. Some concluding remarks are given in Section \ref{sec8}. Proofs are presented in the online Supplementary Material document.

\section{Problem Setup} 
\label{sec2}

{\red 
Let $(\bx_i,y_i)$ be the measurements of the auxiliary variables $\bX$ and the study variable $Y$ associated with unit $i$, $i=1,\cdots,N$, where $N$ is the population size. Let $\theta_N = N^{-1}\sum_{i=1}^Ny_i$ be the finite population mean of the study variable, which is the parameter of interest. 
Let $A$ be a probability sample with observations on the auxiliary variables $\bX$; let $B$ be the non-probability sample with information on both the study variable $Y$ and the auxiliary variables $\bX$. 
Let $n_{A}=|A|$ and $n_{B}=|B|$ be the respective sample sizes. Table \ref{table1} presents the general setup of the two sample structure for data integration. The non-probability sample $B$  is not representative of the target population due to the unknown sample selection/inclusion mechanism. 

\begin{table}[h!]
\centering
\caption{Data Structure for the Two Survey Samples}
\label{table:1}\par
\begin{tabular}{c|c|ccc}
\hline
$\;\;\;$ Sample $\;\;\;$ & $\;\;\;$ Type $\;\;\;$ & $\;\;\;\;$ $\bX$ $\;\;\;\;$ & $\;\;\;\;$ $Y$ $\;\;\;\;$ & $\;$ Representativeness $\;$ \\
\hline
$A$ & Probability Sample & \Checkmark & \xmark & Yes \\
$B$ & Non-probability Sample & \Checkmark & \Checkmark & No \\
\hline 
\end{tabular}
\label{table1} 
\end{table}

The two sample datasets can also be represented by $\{(\bx_i,w_i), i \in A\}$ and $\{(\bx_i,y_i),i\in B\}$, where $w_i=\pi_i^{-1}$ are the survey weights for the probability sample $A$ and $\pi_i = P(i\in A)$ are the first-order inclusion probabilities.
If the study variable $Y$ were observed for the probability sample $A$, the Horvitz-Thompson estimator of $\theta_N$ would be given by $\hat{\theta}_{HT} = N^{-1}\sum_{i\in A}w_iy_i$. The mass imputation estimator to be introduced in the next section replaces the unobserved $y_i$ by an imputed $\hat{y}_i$ using data from both samples $A$ and $B$. 
}

{\red 
Our discussions on important practical issues related to the mass imputation estimator are motivated by the analysis of the survey sample collected by the Pew Research Centre in 2015. 
}
The dataset, denoted by PRC, contains  56 variables which aim to reveal the relations between people and their 
communities. There is a total of 9,301 cases in the PRC dataset which  are provided by eight different vendors with unknown sampling and data collection strategies. We treat the PRC dataset as a non-probability survey sample with the sample size $n_B=9,301$.
{\red 
Four variables of the PRC dataset are of particular interest to our analysis, which are pertinent to the research questions on ``people and their communities'', and are treated as study variables. They are listed at the bottom of Table \ref{table2}, among them three are binary variables: Talk with neighbors frequently ($Y_1$), Participated in school groups ($Y_2$), Participated in service organizations ($Y_3$), and one is treated as a continuous  variable: Days had at least one drink last month ($Y_4$).
}

\begin{table}[!htbp]
\centering
\caption{Estimated Population Means of Survey Items  from the Three Samples}
\label{table2}\par
\begin{tabular}{l|l|ccc}
\hline
 &  & PRC & CPS  & BRFSS \\
  \hline
   Age category & $<$30 & 0.183  & 0.212& 0.209\\
  & $>$=30, $<$50 & 0.326  & 0.336&0.333\\
  & $>$=50, $<$70 & 0.387  & 0.326& 0.327 \\
  & $>$=70 & 0.104 &0.126 & 0.131 \\
   \hline
 Gender & Female & 0.544 &0.518& 0.513  \\
   \hline
    Race&  White only & 0.823  &0.786& 0.750\\
 & Black only&  0.088  & 0.125& 0.126\\
   \hline
   Origin& Hispanic/Latino &0.093  &0.156& 0.165\\
   \hline
      Region & Northeast & 0.200 &0.180& 0.177\\
 
    & South & 0.275 &0.373& 0.383\\
 
    & West & 0.299 &0.235 & 0.232\\
   \hline
   Marital status& Married &   0.503 &0.528& 0.508\\
         \hline
  Employment & Working &  0.521   & 0.589& 0.566\\
 
   & Retired & 0.243  & 0.143& 0.179\\
     \hline
      Education& High school or less & 0.216 & 0.407& 0.427 \\   
      & Bachelor's degree and above &0.416  & 0.309& 0.263  \\
      \hline
  Household & Presence of child in household&  0.289  &NA& 0.368\\
    & Home ownership & 0.654  &NA& 0.672\\
   \hline
       Health& Smoke everyday &0.157 &NA& 0.115\\
   & Smoke never & 0.798   &NA& 0.833\\
   \hline
    Financial status& No money to see doctors & 0.207  &NA& 0.133\\
    & Having medical insurance & 0.891 &NA & 0.878\\
 & Household income $<$ 20K & 0.161 &0.153&NA\\
  & Household income $>$100K & 0.199 & 0.233& NA \\
   \hline
    Volunteer works & Volunteered & 0.510 &0.248& NA\\
   \hline
  \multirow{4}{*}{Study variables}  & Talk with neighbors frequently &0.461    & NA& NA \\
   & Participated in school groups &0.210  & NA & NA \\
& Participated in service organizations &0.141 & NA & NA \\
  & Days had at least one drink last month &5.301 & NA & NA \\
     \hline 
    \end{tabular}
\vspace{1ex}
{\raggedright ``NA'' indicates that the variable is not measured in the survey. \par}
\end{table}

{\red 
The general framework for mass imputation requires the availability of an existing probability sample taken from the same target population and containing a set of auxiliary variables which are common to the non-probability sample. It turns out that there are (at least) two such probability survey samples available, both were taken in 2015 from the general US population. This raises an important practical question on how mass imputation can be implemented under such scenarios, which will be further discussed in Sections \ref{sec7} and \ref{sec8}. 
}

The first probability sample is the volunteer supplement survey data from the Current Population Survey (CPS), which is one of the most  recognized surveys in the United States. The CPS dataset contains $n_A = 80,075$ cases with measurements on volunteerism, which is highly relevant to the study variables considered in the PRC dataset. The second probability  sample is the Behavioral Risk Factor Surveillance System (BRFSS) survey data. It  is  designed to measure behavioral risk factors for US residents and has a large sample size $n_A = 441,456$.  
Neither of the two probability  samples contains  measurements of the study variables, but both share a rich set of common survey items with the PRC dataset as shown in Table \ref{table2}.

{\red 
As a preliminary analysis, we examine marginal distributions of the variables contained in the three datasets.} 
Table \ref{table2} contains the estimated population mean using each of the three datasets. For the PRC dataset, the sampling strategy is unknown and no survey weights are  available, so the estimates presented are the unadjusted simple sample means. For the BRFSS and the CPS datasets where survey weights are available, survey weighed estimates are computed.
The ``NA'' in the table indicates that the variable is not available from the dataset. It can be seen that there are noticeable differences between the naive sample means from the PRC sample and the survey weighted estimates from the two probability samples for variables such as Origin (Hispanic/Latino), Education (High school or less), Household (with children), Health (Smoking) and Volunteer works. It is clear evidence that the PRC dataset is not a representative sample for the US population.

\section{\red The Mass Imputation Estimator}
\label{sec3}

{\red 
We now formally define the mass imputation estimator of the population mean $\theta_N$ under the two-sample setup presented in Table \ref{table2}.
Let $\delta$ be the indicator variable for the unit being included in the non-probability sample $B$. Note that $\delta = 1$ for all $i\in B$ and $\delta = 0$ for all $i \notin B$. This is fundamentally different from the standard setting for missing data analysis where the response indicator variable equals to 1 for respondents and 0 for nonrespondents, all part of the sample. It is also similar to missing data problems if we treat the entire finite population as a sample, which intuitively justifies the need for auxiliary information at the population level in analyzing non-probability survey samples.  
}

We first assume that each unit in the population has a non-zero probability to be included in the sample $B$, i.e.,
\begin{equation}
P ( \delta =1 \mid \bX=\bx) >0
\label{a2}
\end{equation}
for all $\bx$ in the support of $\bX$. 
{\red 
Condition (\ref{a2}) is often referred as as the {\em Positivity  Assumption}, which means that for any possible value $\bx$, there is a positive probability for this type of units to be selected for sample $B$. It implies that the support of $\bX$ in sample $B$ coincides with the sample space of $\bX$ in the population. The condition can easily be verified if all components of $\bX$ are discrete. 
%If assumption (\ref{a2}) is not satisfied, then we do not necessarily have  (\ref{a3})  for all values of $\bX$ in the support of $\bX$. 
If certain components of $\bX$ are continuous, it is still plausible
to check the condition based on the distributions 
of $\bX$ between the weighted distribution in the probability sample $A$ and the empirical distribution in the non-probability sample $B$. If the two sample supports overlap to each other, then (\ref{a2}) is satisfied. \cite{crump2009} proposed a systematic way of  subsampling to handle the cases with limited overlaps in the context of causal inference.   
}

{\red 
The most crucial part of the mass imputation approach is the prediction model $f(y\mid \bx)$, which can be estimated by using the observed $(Y,\bX)$ from the non-probability sample $B$ if 
\begin{equation}
f( y\mid \bx, \delta=1) = f ( y \mid \bx) \,.
\label{a3}
\end{equation}
The prediction model $f(y\mid \bx)$ can then be used to create mass-imputed values of $y$ using observed $\bx$ for all the units in the probability sample $A$. 
}
Equation (\ref{a3}) can be termed as the {\red \emph{transportability} condition} in the sense that the imputation model built from sample $B$ is transportable to sample $A$. 
{\red A sufficient condition for (\ref{a3}) to hold is the ignorability condition for the non-probability sample $B$, which is similar to the missing at random (MAR) assumption  \citep{rubin1976} widely used in the literature on missing data analysis:
\begin{equation}
P( \delta = 1 \mid \bX, Y) = P( \delta =1 \mid \bX) \,.
\label{a1}
\end{equation}
In other words, transportability is a weaker condition than ignorability. Unfortunately, the ignorability condition cannot be formally tested using the sample itself. 
}
%%Assumption (\ref{a1}) is a strong assumption as it means that the sampling mechanism for sample B does not depend on $Y$ after conditioning on $\bX$. Given the data structure in Table \ref{table2}, there is no way to test this assumption. 
%Instead of naively assuming (\ref{a1}), we shall investigate the effect on the mass imputation estimator when assumption (\ref{a1}) is violated. 

%Assumption (\ref{a2}) implies that the sample support of $\bX$ in sample B coincides with the support of $\bX$ in the population. If assumption (\ref{a2}) is not satisfied, then we do not necessarily have  (\ref{a3})  for all values of $\bX$ in the support of $\bX$.

%Unlike the setup of \cite{kim2012combining}, the proposed estimator is no longer design-consistent, but is still justified under the design-model framework, where the model refers to the superpopulation model corresponding to $f(Y|X)$.

Under assumptions (\ref{a2})-(\ref{a3}), it is possible to
consider a mass imputation estimator based on nearest neighbor imputation as suggested by \cite{rivers2007}. Nearest neighbor imputation is a nonparametric method that does not require any parametric model assumptions. While nonparametric imputation methods can provide robust estimation, it suffers from curse of dimensionality, and the asymptotic bias of the nearest neighbor imputation is not negligible if  the dimension of $
\bx$ is greater than one \citep{yang2018}.
In this paper, we consider a semi-parametric model for sample $B$ with the first {\red conditional moment} specified as
\begin{equation}
\label{5.1}
E(Y \mid \bX = \bx)=m(\bx;\bbeta_0)
\end{equation}
for some unknown $p\times 1$ vector $\bbeta_0$ and a known function $m(\cdot;\cdot)$.  
{\red 
The specification of $m(\bx;\bbeta)$ typically follows the mean function for generalized linear models. If $Y$ is continuous, we can use a linear regression model and let $m(\bx;\bbeta) = \bx'\bbeta$; if $Y$ is binary, we can use a logistic regression model and let $m(\bx;\bbeta) =  \{1+\exp(-\bx'\bbeta) \}^{-1}$; if $Y$ represents count data, we can use a log-linear model and let $m(\bx;\bbeta) = \exp(\bx'\bbeta)$. 
Under the assumed transportability condition, the model parameters $\bbeta$ can be estimated by the non-probability sample, $B$. 
}
We assume that $\hat{\bbeta}$ is the unique solution to the estimating equations 
\begin{equation}
\hat{U} ( \bbeta) = \frac{1}{n_B} \sum_{i \in B}\left\lbrace y_{i}-m(\bx_{i};\bbeta) \right\rbrace \bh(\bx_{i};\bbeta)=0
 \label{5.1b}
 \end{equation}
for some \textit{p}-dimensional vector of functions $\bh(\bx;\bbeta)$. 
%We assume that $\bh=\bh( \bx ; \bbeta)$ includes an intercept term such that $\bh' = (1, \bh_1')$.
{\red 
The estimating equations given in (\ref{5.1b}) may be constructed as the quasi-score equations from generalized linear models.  
}
Note that $E\{ \hat{U} ( \bbeta_0) \}=0$ under the assumption (\ref{a3}) and the model (\ref{5.1}).
{\red 
Under suitable moment conditions on $(Y, \bX)$ and smoothness conditions on $m(\bx;\bbeta)$ with respect to $\bbeta$ similar to those discussed in Section 3.2 of \cite{Tsiatis2006}, it can be shown that the solution to (\ref{5.1b}), denoted as $\hat{\bbeta}$, is consistent for $\bbeta_0$ under the ignorability assumption.

Once  $\hat{\bbeta}$ is obtained from the non-probability sample $B$  where both $Y$ and $\bX$ are observed, we can obtain predicted values $\hat{y}_i = m( \mathbf{x}_i ; \hat{\bbeta})$ for all $i \in A$, which is the so-called ``mass imputation''.
The mass imputation estimator for the finite population mean $\theta_N=N^{-1} \sum_{i=1}^N y_i$ is computed as
\begin{equation}
\label{5.2}
\hat{\theta}_{I}=\frac{1}{N} \sum_{i \in A}w_{i}\hat{y}_{i} 
\end{equation}
using the survey weights $w_i$ for the probability sample $A$. If the population size $N$ is unknown, it can be estimated by $\hat{N} = \sum_{i\in A} w_i$. Some asymptotic properties of the mass imputation estimator given in (\ref{5.2}) under the assumed semiparametric model (\ref{5.1})  are presented in the next section.
}

%We also need some smoothness assumptions (such as differentiability with respect to $\beta$) to develop an asymptotic theory for the mass imputation estimator.

%\renewcommand{\theequation}{4.\arabic{equation}}
%\setcounter{equation}{0}

\section{\red  Main Theory}
\label{sec4}

We now discuss asymptotic properties of the mass imputation estimator given in (\ref{5.2}). {\red  \cite{kim2012} investigated the asymptotic properties of the mass imputation estimator in the context of combining two independent probability samples. We now consider an extension to mass imputation using a non-probability sample as a training sample. }
For the asymptotic framework, we assume there is a sequence of finite populations and a sequence of samples as discussed in \cite{fuller2009}, {\red which provides a framework to allow the sample sizes go to infinity. Let $\hat{\bbeta}$ be the solution to (\ref{5.1b}) and define $\bbeta^* = p\lim \hat{\bbeta}$, where the reference distribution is the sampling mechanism for sample $B$. 
Roughly speaking, if (\ref{a1}) holds, then $\bbeta^*= \bbeta_0$, where   $\bbeta_0$ are the true parameters in the conditional model  (\ref{5.1}).
Under certain regularity conditions including   
$ V \{ \hat{U} ( \bbeta^*) \} =O_p (n_B^{-1} ) ,$ 
we can establish that 
\begin{equation} 
\hat{\bbeta} - \bbeta^* = O_p (n_B^{-1/2} ). 
\label{7} 
\end{equation} 
  The following theorem presents an asymptotic expression of the mass imputation estimator. }
%Note that we do not require 

%theorem establishes the consistency and asymptotic variance formula of the mass imputation estimator under the joint randomization of the probability sampling design for sample A and the prediction model for sample B.

%Such assumption can be called transportability in the context of data integration.
%If the variance function $V(y_{i}|\bx_{i})$ is specified as $V(y_i|\bx_i)=\sigma^{2}a(\bx_{i},\bbeta)$ for some known function $a(\cdot)$, then $h(\bx_{i};\bbeta)=\dot{m}(\bx_{i};\bbeta)/a(\bx_{i},\bbeta)$ where $\dot{m}(\bx_{i};\bbeta)=\partial (\bx_{i};\bbeta) / \partial \bbeta$.

\begin{theorem} Suppose that $(\bx,y)$ has bounded fourth moments over the sequence of finite populations  {\red and that (\ref{7}) holds. Under the regularity conditions stated in the Supplementary Material},  
the mass imputation estimator (\ref{5.2}) satisfies
$\hat{\theta}_{I}=\tilde{\theta}_{I}+o_{p}\big(n_{B}^{-1/2}\big)$,
where
\begin{equation}
\label{5.6}
\tilde{\theta}_{I}=N^{-1}\sum_{i \in A}w_{i}m(\bx_{i};\bbeta^*)+N^{-1}\sum_{i \in B}\left\lbrace y_{i}-m(\bx_{i};\bbeta^*) \right\rbrace g(\bx_i; \bbeta^*) 
\end{equation} 
with 
$g(\bx_i; \bbeta^*) =  \bh(\bx_{i};\bbeta^*)' \bc   $,  
and  $\bc$ is the   solution to 
\begin{equation}
\label{5.7}
 \sum_{i \in B}\dot{\bm}(\bx_{i};\bbeta^*) \bh(\bx_{i};\bbeta^*)'\bc  =  \sum_{i=1}^{N}\dot{\bm}(\bx_{i};\bbeta^*)\,,
\end{equation}
where \ $\dot{\bm}(\bx;\bbeta) = \partial m(\bx;\bbeta)/\partial \bbeta$. 
\label{thm1} 
\end{theorem}

{\red 
Proof of Theorem \ref{thm1} is presented in the online Supplementary Material.  
The expression (\ref{5.6}) is essentially the first-order Taylor linearization of $\hat{\theta}_I$ in (\ref{5.2}) and the Taylor expansion is made around $\bbeta = \bbeta^*$, not around the true parameter value $\bbeta_0$. Note that we do not use  the ignorability  assumption in (\ref{a1}) to establish (\ref{5.6}).

Following (\ref{5.6}), we can express 
\begin{equation} 
 \tilde{\theta}_I - \theta_N = N^{-1}\left( \sum_{i \in A}w_{i}m_i^* -  \sum_{i=1}^N m_i^* \right) +  N^{-1} \left(   \sum_{i =1}^N  \delta_i g_i^*e_i^*   -  \sum_{i=1}^N e_i^*  \right) \,,
\label{5.6b}
\end{equation}
where $m_i^* = m(\bx_i; \bbeta^*)$,  $e_i^* = y_i - m(\bx_i; \bbeta^*)$ and $g_i^* = g(\bx_i; \bbeta^*)$. 
Thus, ignoring the smaller order terms, the asymptotic bias 
is 
\begin{eqnarray}
E( \tilde{\theta}_I - \theta_N)   &=& E\left\{ N^{-1} \left(   \sum_{i =1}^N  \delta_i g_i^*e_i^*   -  \sum_{i=1}^N e_i^*  \right) \right\}   \notag \\
&=& -   E\left(  N^{-1}   \sum_{i=1}^N e_i^*  \right) \notag \\
&=&  E\left\{ Cov_N ( \delta_i \pi_B^{-1} , e_i^* )  \right\}\,, \label{bias1} 
 \end{eqnarray} 
where $\pi_B = n_B/N$ and the second equality follows from $ E( \delta_i e_i^* \mid \bx_i) =0$  by the definition of $\bbeta^*$ and $e_i^*$.  Here, we used the notation  $Cov_N (x_i, y_i) = N^{-1} \sum_{i=1}^N (x_i - \bar{x}_N) (y_i - \bar{y}_N)$. 
If  $\bbeta^* = \bbeta_0$, then $e_i^*=y_i - m(\bx_i; \bbeta_0)= e_i$ and 
the mass imputation estimator in (\ref{5.2}) is unbiased. 
Otherwise, the bias is non-zero. 
%The sign of the bias depends on the partial correlation between $\delta_i$ and $y_i$ after conditioning on $x_i$. 

On the other hand, the asymptotic bias of the naive estimator, the simple sample mean $\bar{y}_B = n_B^{-1} \sum_{i \in B} y_i$,  is given by 
\begin{eqnarray} E( \bar{y}_B - \theta_N )  &=& E\{ n_B^{-1} \sum_{i =1}^N \delta_i (y_i - \bar{Y}_N)  \} \notag  \\&=& E\{ Cov_N (  \delta_i \pi_B^{-1}  , y_i) \} .  \label{bias2} \end{eqnarray}  
 Thus, comparing (\ref{bias1}) with (\ref{bias2}), we find that the absolute value of the bias of $\tilde{\theta}_I$ is smaller than that of $\bar{y}_B$ as the variance of $e_i^*$ will be smaller than the variance of $y_i$. Thus, the mass imputation estimator reduces the bias even when the sampling mechanism for sample $B$ is non-ignorable. 
 
 We now derive the asymptotic variance formula for the mass imputation estimator. By (\ref{5.6b}), 
 we have 
$V(\tilde{\theta}_{I}-\theta_{N}) = V_A + V_B$, where
\begin{equation}
\label{5.8}
V_A =  V\left( N^{-1} \sum_{i \in A} w_i m_i^* - N^{-1} \sum_{i=1}^N m_i^* \right) 
\end{equation}
 is the variance component for sample $A$ and
 \begin{eqnarray}
V_B&=&  V\{  
N^{-1} (   \sum_{i =1}^N  \delta_i g_i e_i^*   -  \sum_{i=1}^N e_i^*  ) \}\notag \\
&=& V[  E \{ 
N^{-1}   \sum_{i =1}^N  ( \delta_i g_i -1) e_i^*   \mid \bX,  \bdelta \}   ] + E [  V \{ 
N^{-1}    \sum_{i =1}^N  (\delta_i g_i-1) e_i^* \mid \bX, \bdelta \}  ] 
\label{5.10}
\end{eqnarray} 
is the variance component for sample $B$. In (\ref{5.8}), only the design-based variance under the probability sampling design for sample $A$ remains. 
The reference distribution in (\ref{5.10}) is the joint distribution of the sampling mechanism for sample $B$ and the  superpopulation model in (\ref{5.1}). The conditional expectation $E\{ \cdot \mid \bX,  \bdelta\}$ is with respect  to the model for  $e_i^* = y_i - m( \bx_i; \bbeta^*)$ conditional on $\bdelta$. The first term of (\ref{5.10}) is unknown in general because the sampling mechanism for sample $B$ is unknown.

If the sampling mechanism for sample $B$ is ignorable as defined in (\ref{a1}), we have $\bbeta^* = \bbeta_0$ and $e_i^* = e_i$. In this case, the first term of (\ref{5.10}) disappears  because $E ( e_i \mid \bx_i, \delta_i ) = E\{ y_i - m( \bx_i; \bbeta_0) \mid \bx_i, \delta_i \}=E\{ y_i - m( \bx_i; \bbeta_0) \mid \bx_i \}=0$. Thus, under the ignorability assumption, we can simplify $V_B$ as 
 \begin{eqnarray}
V_B&=&   E \left[  V \left\{ 
N^{-1}    \sum_{i =1}^N  (\delta_i g_i-1) e_i \mid \bX, \bdelta \right\}  \right] \notag \\
&=& E\left\{ N^{-2} \sum_{i=1}^N (\delta_i g_i-1)^2 e_i^2 \right\} .  
\label{5.9}
\end{eqnarray} 
The second equality follows by the independence of $e_i$'s in the superpopulation model. 
Thus, the first term of (\ref{5.10}) can be understood as the additional variance increase under  model misspecification. 
Furthermore, if $n_B/N=o(1)$ then we can simply use
\begin{equation}
\label{5.9b}
V_B= E \bigg[ N^{-2}\sum_{i \in B} \left(  e_i g_i \right)^2 \bigg].
\end{equation}
}

Note that we can easily estimate $V_B$ in (\ref{5.9b}) even though the sampling mechanism for sample $B$ is unknown.
%Proof of Theorem 1 is presented in Appendix A. 
The asymptotic variance $V(\tilde{\theta}_{I}-\theta_{N})$ consists of two parts. The first term $V_A$ is of order $O(n_A^{-1})$ and the second term $V_B$  is of order $O(n_B^{-1})$. If $n_{A}/n_{B}=o(1)$, i.e., the sample size $n_B$ is much larger than $n_A$, the term $V_{B}$ is of smaller order and the leading term of the total variance is $V_{A}$. Otherwise the two variance components both contribute to the total variance. 
%In big data applications where $n_{B}$ is very large, the term $V_{B}$ can be safely ignored.

Under the linear regression model $Y_i = \mathbf{x}_i' \bbeta + e_i $ with $e_i \sim (0, \sigma_{e}^{2})$, independent among all $i$, we have  $\hat{y}_i = \mathbf{x}_i' \hat{\bbeta}$ where
$ \hat{\bbeta} = \left( \sum_{i \in B} \mathbf{x}_i \mathbf{x}_i' \right)^{-1}  \sum_{i \in B} \mathbf{x}_i y_i $.
The mass imputation estimator of (\ref{5.2}) under the regression model is given by $\hat{\theta}_{I,reg} = N^{-1}\sum_{i\in A}w_i\bx_i'\hat\bbeta$.
If the probability sample $A$ is selected by simple random sampling, the asymptotic variance of $\hat{\theta}_{I,reg}$ is given by
\begin{equation}
V \left( \hat{\theta}_{I, reg} - \theta_N \right) \approx V \bigg( n_A^{-1} \sum_{i \in A} \mathbf{x}_i' \bbeta \bigg)   + V \bigg( N^{-1} \sum_{i \in B} e_i \bx_i' \bc  \bigg) \,,
\label{v2}
\end{equation}
where
$ \bc = \left( N^{-1} \sum_{i \in B} \mathbf{x}_i \mathbf{x}_i' \right)^{-1} \bar{\bx}_N$ and $\bar{\bx}_N = N^{-1} \sum_{i=1}^N \mathbf{x}_i$.
If $\mathbf{x}_i = (1, x_i)'$ and $\bbeta = (\beta_0,\beta_1)'$ and assuming $\pi_B = n_B/N$ is negligible,  the asymptotic variance reduces to
\begin{eqnarray*}
V \left( \hat{\theta}_{I, reg} - \theta_N \right) &\approx&  \frac{1}{n_A} (\beta_{1} )^{2}\sigma_{x}^{2}+ \frac{1}{n_{B}} \sigma_{e}^2  +  E\left[\frac{ (\bar{x}_N -\bar{x}_{B})^2 } {\sum_{i \in B}(x_{i}-\bar{x}_{B})^{2}}\right]  \sigma_{e}^{2}  \,,
\end{eqnarray*}
where $\sigma_x^2 = V(X)$  and $\bar{x}_B = n_B^{-1}\sum_{i\in B}x_i$.  If sample $B$ is a random sample from the population, then the third term is of order $O(n_B^{-2})$ and becomes negligible. However, since sample $B$ is a non-probability sample, the third term might not be negligible.

%\section{Variance estimation}

Variance estimation for the mass imputation estimator (\ref{5.2}) requires estimation of the two components $V_A$ and $V_B$. The first component can be estimated by
%If $\bbeta_0$ is known, then $m(\bx_{i};\bbeta_0)$ is observable for all $i \in A$ and we can use the design-unbiased variance estimator, that is
\begin{equation*}
\hat{V}_{A}=\frac{1}{N^2} \sum_{i \in A}\sum_{j \in A}  \frac{\pi_{ij}-\pi_{i}\pi_{j}}{\pi_{ij}} w_{i}m(\bx_{i};\hat{\bbeta})w_{j}m(\bx_{j};\hat{\bbeta}) \,,
\end{equation*}
where  $\pi_{ij} = P(i,j\in A)$ are the joint inclusion probabilities and are assumed to be positive.
%Note that we can easily estimate $V_B$ in (\ref{5.9b}) even though the sampling mechanism for sample B is unknown.
The second component can be estimated by
\begin{equation}
\hat{V}_{B}= \frac{1}{N^2} \sum_{i \in B} \hat{e}_{i}^{2} \big\{\hat{\bc}' \bh(\bx_{i};\hat{\bbeta})\big\}^2 \,,
\label{vb}
\end{equation}
where $\hat{e}_{i}=y_{i}-m(\bx_{i};\hat{\bbeta})$ and
$\hat{\bc} =  \left( N^{-1}  \sum_{i \in B} \mathbf{x}_i \mathbf{x}_i' \right)^{-1} N^{-1} \sum_{i \in A} w_i  \mathbf{x}_i$.
The total variance of $\hat{\theta}_{I}$ can be estimated by $\hat{V} ( \hat{\theta}_I - \theta_N ) = \hat{V}_{A}+\hat{V}_{B}$.

%If $n_{A}/n_{B}=o(1)$, then $V_{B}$ is smaller order than $V_{A}$ and total variance is dominated by $V_{A}$. Otherwise, the two variances both contribute to the total variance. In the big data application, $n_{B}$ is huge and $V_{B}$ can be safely ignored.

%\renewcommand{\theequation}{5.\arabic{equation}}
%\setcounter{equation}{0}

\section{Bootstrap Variance Estimation}
\label{sec5}

The variance estimator presented in Section 4 is based on the linearization method. The closed-form formula for the asymptotic variance is simple to implement. However, to compute $\hat{V}_B$ in (\ref{vb}), we need to use individual observations of $(\bx_i, y_i)$ in sample $B$, which is not necessarily available when only the sample $A$ with mass imputed responses is released to the public data users.
Note that one of the goals of mass imputation is to produce a representative sample $A$ with synthetic observations on the response variable using sample $B$ as a training dataset. Once the mass imputation is performed, the training data is no longer necessary in computing point estimators. It is therefore desirable to develop a variance estimation method that does not require access to observations in sample $B$.

To achieve this goal, we propose a bootstrap method \citep{raowu88} for variance estimation that creates a replicated set of synthetic data $\{ \hat{y}_i^{(k)},
 i  \in A \}$ corresponding to each set of bootstrap weights $\{ w^{(k)}_{i} , i  \in A \}$, $k=1,\cdots,L$ associated with sample $A$ only. The method enables users to correctly estimate the variance of the mass imputation estimator $\hat{\theta}_I$ without access to the training data $\{(y_i , \bx_i ) : i \in B \}$ from sample $B$. The data file will contain additional columns of $\{ \hat{y}^{(k)}_
i : i \in A\}$ associated with the columns of  bootstrap weights $\{w^{(k)}_i ;  i  \in A \}$, $k =1, \cdots , L$, where $L$ is the number of replicates
created from sample $A$. 
\cite{kim2012} also considered a similar method in the context of survey integration from non-nested two-phase sampling.

In order to develop a valid bootstrap method for the mass imputation estimator $\hat{\theta}_I$ in (\ref{5.2}), it is critical to develop a valid bootstrap method for estimating $V( \hat{\bbeta} ) $ when $\hat{\bbeta}$ is computed from (\ref{5.1b}).  Note that, under assumptions (\ref{a3}) and (\ref{5.1}), we can obtain
\begin{equation}
V ( \hat{\bbeta} ) \doteq  J^{-1}  \Omega J^{-1'}
\label{var2}
\end{equation}
where $ J =  E\left\{ n_B^{-1} \sum_{i \in B} \dot{\bm}_i \bh_i'  \right\} $, \
$ \Omega = E\left\{ n_B^{-2} \sum_{i \in B} E( e_i^2 \mid \bx)  \bh_i \bh_i' \right\} $
with $\dot{\bm}_i = \dot{\bm} ( \mathbf{x}_i; \bbeta_0 )$ and $\bh_i = \bh ( \bx_i; \bbeta_0)$. The reference distribution for (\ref{var2}) is the joint distribution of the superpopulation model (\ref{5.1}) and the unknown sampling mechanism for the non-probability sample $B$. Interestingly, the variance formula in (\ref{var2}) equals  exactly to the variance of $\hat{\bbeta}$ when  sample $B$ is selected by simple random sampling (SRS). That is, even though the sampling design for sample $B$ is not SRS, its effect on the variance of $\hat{\bbeta}$ is essentially the same with SRS. This is due to the MAR assumption in (\ref{a3}) which makes the effect of the sampling design for estimating $\bbeta$ ignorable even though it is still not ignorable for $\theta= E(Y)$. Therefore, we can safely ignore the sampling design for sample $B$ when estimating $\bbeta$ and develop a valid bootstrap method for variance estimation of $\hat{\bbeta}$ using the bootstrap method for SRS.

Our proposed bootstrap method can be described as the following four steps:
\begin{itemize}
\item[Step] 1. Create the $k$th set of replication weights $\{w_i^{(k)}, i\in A\}$ based on the sampling design for the probability sample $A$.
\item[Step] 2. Generate the $k$th bootstrap sample of size $n_B$ from sample $B$ using simple random sampling with replacement and compute $\hat{\bbeta}^{(k)}$ using the same estimation equations (\ref{5.1b}) applied to the bootstrap sample.
\item[Step] 3. Use $\hat{\bbeta}^{(k)}$ obtained from Step 2 to compute $\hat{y}_i^{(k)} = m( \bx_i; \hat{\bbeta}^{(k)})$ for each $i \in A$ and create a new column $\{\hat{y}_i^{(k)},i\in A\}$ alongside $\{w_i^{(k)}, i\in A\}$ for the sample $A$ dataset.
\item[Step] 4. Repeat Steps 1-3, independently, for $k=1,\cdots, L$ for a pre-chosen $L$.
 \end{itemize}

 Step 1 needs to follow standard practice in survey sampling on creating replication weights for design-based variance estimation. 
 {\red 
 See, for instance, Chapter 10 of \cite{WuThompson2020}. 
 }
 The final sample $A$ dataset contains additional columns for the replicate versions of mass imputed values $\{\hat{y}_i^{(k)},i\in A\}$ and the sets of replication weights $\{w_i^{(k)}, i\in A\}$, $k=1,\cdots,L$.  The replicate versions of the mass imputation estimator $\hat{\theta}_I$ are computed as
\begin{equation}
 \hat{\theta}_I^{(k)} =\frac{1}{N}\sum_{i \in A} w_i^{(k)} \hat{y}_i^{(k)} \,, \;\;\;\;\;\; k=1,\cdots,L\,,
 \label{14}
 \end{equation}
and the resulting bootstrap variance estimator of $\hat{\theta}_I$  is computed as
\begin{equation}
\hat{V}_{b} ( \hat{\theta}_I ) = \frac{1}{L} \sum_{k=1}^L \left( \hat{\theta}_I^{(k)} - \hat{\theta}_I \right)^2.
\label{bt}
\end{equation}
{Note that once $\{ (w^{(k)}_i ,   \hat{y}^{(k)}_
i ): i \in A\}$ are obtained for $k=1, \cdots, L$, in addition to the original mass imputation data $\{ (w_i ,   \hat{y}_
i ): i \in A\}$,  the users do not need to have access to sample $B$. 
}

The following theorem establishes the consistency of the proposed bootstrap variance estimator.
Proof of the theorem is presented in the online Supplementary  Material.

\begin{theorem}
Suppose that the assumptions of Theorem 1 hold. Assume that the sampling mechanism for sample B is ignorable as defined in (\ref{a1}). 
Under the additional assumption stated in the Supplementary Material document, the bootstrap variance estimator given by (\ref{bt})
satisfies
\begin{equation}
\hat{V}_{b}\big(\hat{\theta}_{I}\big)=V\big(\hat{\theta}_{I} - \theta_N\big)+o_{p}\big(n_{B}^{-1}\big) \,.
\end{equation}
\end{theorem}

\section{Simulation Study}
\label{sec6}

{\red 
A limited simulation study is performed to (i) evaluate the departure of the imputation model to the performance of mass imputation estimator, (ii) check the validity of the proposed variance estimators, and (iii) compare the mass imputation estimator with the inverse probability weighted (IPW) estimator of \cite{chen2018doubly} using a logistic regression model for estimating the propensity scores.

 The setup for the simulation study employed  a $3 \times  2$ factorial structure with two factors. The first factor is the superpopulation model that generates the finite population. The second factor is  the sample size  for sample B.  We used two levels for $n_B$, where  $n_B=500$ or $n_B=1,000$. }
 We generated
 the following three models for finite  populations of size $N=100,000$. {\red The variables, $x_i$ and $e_i$, are independently generated from $N(2,1)$ and $N(0,1)$, respectively. 
 
 The study variable $y_i$ are constructed differently for each model: 
 \begin{enumerate}
 \item[\rm] Model I: \ \ \ $y_i = 1 +2x_i + e_i$. 
 \item[\rm] Model II: \ \ $y_i = 3 + x_i + 2e_i$. 
 \item[\rm] Model III: \  $y_i = 2.5 + 0.5x_i^2 + e_i$. 
 \end{enumerate}
 Model I generates a finite population with a high correlation between $x$ and $y$ ($r^2 = 0.8)$, Model II generates a finite population with a low correlation ($r^2 = 0.2$), and Model III generates a finite population where the linear relationship fails. Model III is included to check the effect of model mis-specification for the imputation model.
}

From each of the three populations, we generated two independent samples. We use  simple random sampling of size $n_{A}=500$ to obtain  sample $A$. In selecting sample $B$ of size $n_{B}$, where $n_B \in \{500, 1000 \}$, we create two strata where Stratum 1 consists of elements with $x_{i} \leq 2$ and Stratum 2 consists of elements with $x_{i}>2$. Within each stratum, we select $n_{h}$ elements by simple random sampling, independent between the two strata,  where $n_{1}=0.7 n_B$ and $n_{2}=0.3 n_B$. We assume that the stratum information is unavailable at the time of data analysis.
Using the two samples $A$ and $B$, we compute four estimators of $\theta_N = N^{-1} \sum_{i=1}^N y_i$:

\medskip

\no
1) The sample mean  from sample $A$: \  $\hat{\theta}_{A}=n_A^{-1} \sum_{i \in A} y_i$.

\medskip

\no
2) The naive estimator (sample mean) from sample $B$:  \ $\hat{\theta}_{B} = n_B^{-1} \sum_{i \in B} y_i $.

\medskip

\no
3) The mass imputation estimator from sample $A$  given in (\ref{5.2}) using $\hat{y}_i = \hat{\beta}_0 + \hat{\beta}_1 x_i$ where $(\hat{\beta}_0, \hat{\beta}_1)$ are  the estimated regression coefficients obtained from sample $B$.

\medskip

\no
4) The inverse probability weighted (IPW) estimator proposed by \cite{chen2018doubly}:
$\hat{\theta}_{IPW}=N^{-1}\sum_{i \in B}\hat{\pi}_{i}^{-1}y_{i}$,
where the propensity scores, $\pi_{i}=\pi(\bx_{i};\bphi) = \{1+\exp(-\phi_0 - \phi_1 x_i ) \}^{-1}$ with $\bx_i = (1,x_i)'$ and $\bphi = (\phi_0,\phi_1)'$, are estimated by using $\hat{\bphi}$ which solves the following score equations:
\begin{equation}
U(\bphi)=\sum_{i \in B}\bx_{i}-\sum_{i \in A}w_{i}\pi(\bx_{i};\bphi)\bx_{i}=\textbf{0}.
\label{eq1}
\end{equation}

The sample mean of sample $A$, not available in practice but computed  in the simulation, serves as a gold standard estimator. Results are based on $1,000$ repeated simulation runs.
Table \ref{table:2} presents the Monte Carlo bias, the Monte Carlo variance, and the relative mean squared error of the four point estimators. The relative mean squared error  of estimator $\hat{\theta}$ is defined as
\begin{equation*}
ReMSE =\frac{MSE(\hat{\theta})}{MSE(\hat{\theta}_{A})} \,.
\end{equation*}
Note that the sample mean $\hat{\theta}_A$ from sample $A$ is  not available in practice but is computed here as the gold standard. Table \ref{table:2} shows that, with models I and  II where the linear regression model holds, the mass imputation estimator is unbiased for the population mean.  The naive mean estimator of sample $B$ underestimates the population mean for all scenarios considered in the simulation. When the size of sample $B$ for training data is larger than the size of sample $A$ ($n_B=1,000$), it is possible that the mass imputation estimator has a smaller MSE than the gold standard. Under the simple regression model,
the asymptotic  variance of the mass imputation estimator is
\[
V\big(\hat{\theta}_{I} - \theta_N\big) \approx \frac{1}{n_A} \sigma_x^2 \beta_1^2 + \frac{1}{n_B} \sigma_e^2 E  \left\{ 1 + \frac{ (\bar{x}_N - \bar{x}_B)^2}{ s_{x,B}^2 } \right\} \,,
\]
where $s_{x,B}^2 = (n_B-1)^{-1}\sum_{i\in B}(x_i - \bar{x}_B)^2$, while the variance of sample mean of sample A  is $
V(\hat{\theta}_{A})=\sigma_{y}^{2}/n_{A}=(\beta_{1}^{2}\sigma_{x}^{2}+\sigma_{e}^{2})/n_{A} $. Thus, if $n_B$ is much larger than $n_A$, the mass imputation estimator can be more efficient than the sample mean of $A$. The IPW estimator is less efficient than the mass imputation estimator for all cases considered in the current simulation setup.

The imputation model using the simple linear regression is incorrectly specified for model III. The mass imputation estimator is modestly biased. Nonetheless, the performance in terms of MSE is better than the IPW estimator because the mass imputation estimator has much smaller variance than the IPW estimator, even though the absolute bias is larger, when the linear relationship fails.

%where $\mu_{x}$ is a population mean of $\bx$ and $\bar{x}_{B}$ is a sample mean of sample $B$.

%Similarly, the variance of the mass imputation estimator for case 2 with linear regression population in Table \ref{table:4} can be computed by
%\begin{eqnarray*}
%V(\hat{\theta}_{I}) &\cong&
%\mbox{$(n_{A}^{-1} - N^{-1})\beta_{1}^{2}\sigma_{x}^{2}+ [ n_{B}^{-1}+ \{ (\mu_{x}-\bar{x}_{B})^2 + \sigma_{x}^{2}/N \}/\{\sum_{i \in B}(x_{i}-\bar{x}_{B})\}^{2} ]\sigma_{e}^{2}$} \\
%&=&(500^{-1}-50,000^{-1})\times 1.2^2 + \{500^{-1}+(1.3345+50,000^{-1})/521.313\} \\
%&=& 0.0074.
%\end{eqnarray*}

Table \ref{table:3} presents Monte Carlo mean and relative bias of the two variance estimators of the mass imputation estimator using linearization and bootstrap.  Both variance estimators show negligible relative biases. In particular, the bootstrap variance estimator shows good performance even under model III.

\section{Application to the PRC Dataset}
\label{sec7}

We now apply the proposed mass imputation techniques described in Section \ref{sec3} and compute point and variance estimates of population means  for 
{\red 
the four study variables in the PRC dataset, which were introduced in Section 2:
\begin{itemize}
    \item[$Y_1$:] Talk with neighbors frequently, 
    \item[$Y_2$:] Participated in school groups, 
    \item[$Y_3$:] Participated in service organizations, 
    \item[$Y_4$:] Days had at least one drink last month.
\end{itemize}
Note that PRC is the non-probability sample and CPS and BRFSS are the two existing probability samples to be used to provide information on  auxiliary variables. The PRC sample was collected through web panels, the BRFSS survey was conducted by telephone interviews, and the CPS survey used a mixed mode with 11 percent of units surveyed by telephone and the remaining part of the survey by personal interviews. Further details are provided in the online  Supplementary Material document. 

Prior to any formal analysis of the datasets, we would like to check whether the positivity and the transportability assumptions are reasonable. The support of the auxiliary variables in the PRC dataset can be checked and compared to the support in the CPS sample or the BRFSS sample, which confirms the positivity assumption. The transportability assumption cannot be formally tested. We first looked into the auxiliary variables to see if they are meaningful predictors for the study variables, and the answer is clearly affirmative. We then conducted a sensitivity analysis by (i) choosing a ``test study variable'' $Z$ which is available from the non-probability sample and one of the probability samples; (ii) fitting the test model $f(z \mid \bx)$ separately by using the non-probability sample and the probability sample; (iii) comparing the two fitted models for similarities and differences. Details of the sensitivity analysis are reported in the Supplementary Material document. Note that results from such analysis do not formally prove or disprove the validity of the assumption. For instance, the transportability of the model $f(x_2 \mid x_1)$ does not imply the transportability of the model $f(y\mid x_1,x_2)$.
}

Our analysis focuses on three important practical aspects of the method. We first investigate how the relative sizes of sample $A$ and sample $B$ impact the mass imputation estimator and  its associated variance estimator. 
We then discuss the covariates selection for the prediction model, and apply different strategies based on the availability of auxiliary variables from one or both probability samples. Lastly, we make comparisons between the mass imputation estimators and the IPW estimators.

\subsection{Impact of relative sample sizes}

Recall from Section 2 that the two probability samples CPS and BRFSS contain respectively $n_A = 80,075$ and $441,456$ cases while the non-probability PRC  dataset has $n_{B}=9,301$ units. The sample size $n_{A}$  is much larger than $n_{B}$ for both CPS and BRFSS. To investigate other possible scenarios where $n_{A}$  and $n_{B}$ has different ratios, we draw three subsamples from the original BRFSS dataset by the simple random sampling without replacement method. The resulting subsamples, denoted by  BRFSS\textsuperscript{(1)}, BRFSS\textsuperscript{(2)} and BRFSS\textsuperscript{(3)}, have sample size  $n_A^*=80,000$, $8,000$ and $800$, respectively. 
Survey weights for each of the subsamples are computed as $w_i n_{A}/n_{A}^*$, where $w_i$ is the weight  of unit $i$ in the original  BRFSS sample with $n_{A}=441,456$.

The choices of covariates for the prediction model are constrained by the set of covariates observed in both the probability sample and the non-probability sample. The common set of covariates between  BRFSS and PRC,  however, differs from the set between CPS and PRC. We use the set of covariates which are available in all three datasets for the current investigation. We use a logistic regression model for  binary responses and a linear regression model for the continuous response. 

We first compute the point estimates of the population means by the proposed mass imputation method using five different probability samples, and the results are presented in Table \ref{table:6}. The first row specifies which probability sample is used as sample $A$ except for the column under ``PRC'' which represents the naive estimates of simple sample means.
It can be seen that the three larger probability samples BRFSS, BRFSS\textsuperscript{(1)} and BRFSS\textsuperscript{(2)} with $n_{A}=441,456$ and $n_A^*=80,000$  and $8,000$ produce almost identical results. The smallest probability sample BRFSS\textsuperscript{(3)} with $n_A^*=800$ leads to noticeably different results, indicating potential inconsistent estimates when the size of the probability sample is too small. We also notice that, despite the differences between the CPS and the BRFSS samples, their corresponding  mass imputation  estimators  produce similar results 
{\red  with the use of the same set of auxiliary variables}.

We further look at  variance estimators with different probability samples, using the linearization and bootstrap methods described in Sections 4 and 5. The linearization variance estimator  is based on the formula $V\big(  \tilde{\theta}_{I} - \theta_{N}\big) = V_A + V_B$ given by Theorem 1, where $V_A$ is the design-based variance component under the probability sampling design for sample $A$, and  $V_B$ is the variance component for sample $B$ under the prediction model. Unfortunately, detailed design information other than the survey weights is not available for either the BRFSS or the CPS sample. We use an approximate variance formula for $V_A$ by assuming that the survey design is single-stage PPS sampling with replacement, a strategy often used by survey data analyst for the purpose of variance estimation. The bootstrap variance estimator  is computed based on $L=5,000$ bootstrap samples.

Results for variance estimators with decomposition of the variance components ($\hat{V}_A, \hat{V}_B$) for the linearization method are reported in Table \ref{table:7}.  The variance estimates  have been multiplied by $10^5$ for binary response variables and  by $10^2$ for the continuous variable. Variance decomposition cannot be done for the bootstrap method. We have the following major observations from Table \ref{table:7}: (i) the two variance estimation methods, linearization and bootstrap, give similar results in total variance for all cases; (ii) the two original large probability samples CPS and BRFSS produce similar total variances for all cases, with the design variance component $\hat{V}_A$ making a negligible contribution; (iii) when the size of the probability sample becomes smaller, from BRFSS\textsuperscript{(1)} to BRFSS\textsuperscript{(2)} and then BRFSS\textsuperscript{(3)}, the total variance becomes larger, and the variance component $\hat{V}_A$ from the probability sampling design for sample $A$ becomes dominant.

\subsection{Impact of the prediction model} 

The second part of the analysis is on covariate selection for the prediction model. We consider following selection strategies: (a) Use the common set of variables available in all three datasets. This is what has been used in Section 7.1, and the resulting model is denoted as $\xi({\rm Partial})$. (b) Use the common set of variables available in the two datasets PRC and CPS or PRC and BRFSS, leading to two larger but different common sets of variables for the two probability samples. The resulting models are denoted as $\xi({\rm All})$. (c) Use the set of covariates which are available in both PRC and CPS (or both PRC and BRFSS) but are selected through the backward variable selection algorithm. The resulting models are denoted as $\xi({\rm Select})$. P-values for covariates in  model $\xi({\rm All})$ and $\xi({\rm Select})$ are listed in Table \ref{table:8} and Table \ref{table:9}.

Mass imputation based point estimates $\hat{\theta}_{I}$ for the population means of the four response variables using different sets of covariates with the probability Sample $A$ as either CPS or BRFSS are presented in Table \ref{table:10}, with the linearization variance estimates displayed in the parentheses (multiplied by $10^5$ for binary variables and by $10^2$ for the continuous variable). Major observations from Table \ref{table:10} can be summarized as follows. (i) With a chosen sample $A$, the point estimates under model $\xi({\rm All})$ are very similar to the estimates under model $\xi({\rm Select})$ which drops some covariates using the backward variable selection algorithm; (ii) the point estimates under model $\xi({\rm All})$ behave quite differently to the estimates under model $\xi({\rm Partial})$ which excludes some covariates based on the availability from a second probability sample; (iii) the linearization variance estimates under model $\xi({\rm Select})$ are always smaller than the variance estimates under model $\xi({\rm All})$, showing some efficiency gain by eliminating non-significant factors from the model; (iv) the point estimates using CPS are very similar to the estimates using BRFSS under the same model $\xi({\rm Partial})$ with the same set of covariates; (v) the point estimates using CPS under model $\xi({\rm All})$ are quite different to the estimates using BRFSS under model $\xi({\rm All})$. Note that the two models use two different sets of covariates. Some additional covariates available with one particular probability sample may be highly correlated to some of the response variables. 
For instance,  the  covariate ``Volunteer works'', which is  available in CPS but not in BRFSS, is related to the response variables $Y_1$, $Y_2$ and $Y_3$. Similarly, health related covariates, which are available in BRFSS but not in CPS, likely explain the evident difference of $\hat{\theta}_{I}$ for response  $Y_4$ when the BRFSS probability sample is used.

\subsection{Comparing mass imputation with inverse probability weighting} 

In the final part of the analysis of the PRC non-probability sample, we make comparisons between mass imputation estimator $\hat{\theta}_{I}$ and IPW estimator $\hat{\theta}_{IPW}$ of \cite{chen2018doubly}, which is also examined in the simulation study reported in Section 6. The mass imputation estimates are computed based on model $\xi({\rm Select})$. The IPW estimators are based on the logistic regression model for the propensity scores using the same set of covariates as in the model $\xi({\rm Select})$. 
The point estimates along with the associated linearization variance estimates (multiplied by $10^5$ for binary variables and by $10^2$ for the continuous variable) for the four response variables are reported in Table \ref{table:11}. The two estimators $\hat{\theta}_{I}$ and $\hat{\theta}_{IPW}$ provide   comparable results in both bias and variance, regardless whether CPS or BRFSS is used as the probability sample $A$. 

%%We have also analyzed other response variables which are available in the PRC non-probability survey samples, such as ``Tended to trust neighbors'', ``Expressed opinions at a government level'', ``Voted local elections'', ``Participated in sports organizations'', and ``No money to buy food''. The results, not reported here to save space, convey the same messages as those from Tables \ref{table:7}-\ref{table:11}. 

\section{\red Concluding Remarks}
\label{sec8}

The use of non-probability survey samples as an efficient and cost-effective data source has become increasingly popular in recent years. Theoretical developments on analysis of non-probability samples, however, severely lag behind the need of making valid inference from such datasets. Non-probability survey samples are biased and do not represent the target population. Valid inferences require supplementary information on the population. The mass imputation approach proposed here relies on the availability of a probability survey sample from the same target population with information on covariates. The covariates are also measured for the non-probability sample and need to possess two crucial features for the framework discussed in this paper: 
{\red 
(a) they characterize the inclusion/exclusion mechanism for units in the non-probability sample so that the ignorability or the transportability assumption could be justified; and (b) they are relevant to the response variable in terms of prediction power, a condition required for the mass imputation estimator.

Our analyses of the PRC non-probability sample shed light on two important practical issues. The first is the choice of an existing probability sample from the same target population if two or more such samples are available. The decision rests on the set of auxiliary variables which is available in both the non-probability and the probability samples. The size of the probability sample is less crucial as long as it is not too small. Two different probability samples tend to produce similar results if the same set of auxiliary variables is used. The second is the interpretation of the mass imputation estimator when the non-probability survey and the probability survey use different modes of data collection. The estimator is computed using the probability sample under mass imputation but the final estimate carries the mode from the non-probability sample since the model is built based on the observed study variable in the non-probability sample. 
}

{\red  One of the main advantages of the mass imputation approach over the propensity score approach is that we can use  subject matter knowledge on the study variables  to build a good prediction model that can lead to  better estimation.
Also, as discussed in Section 4, the bias of the mass imputation estimator can be small even in misspecified models. Thus,  mass imputation can provide  rather robust estimates. The price we pay is the extra modeling work for each survey item.

The theoretical results on mass imputation presented in this paper focus on the estimation of finite population means. This is in line with traditional approaches in survey sampling where the basic theory is developed for the Horvitz-Thompson estimator. Inferences on other finite population parameters under mass imputation remain to be a research topic for future development. The main theoretical results are also based on deterministic imputation under the semiparametric model (\ref{5.1}). Another research topic is to explore the use of random imputation, including multiple imputation \citep{rubin87} and fractional imputation \citep{kim11}, for analyzing non-probability survey samples.  
}

Statistics Canada has been implementing the modernization initiatives in recent years, which call for a culture switch from the traditional survey-centric approach by the agency to using data from multiple sources. One of the questions arising from the discussions is the role of traditional probability-based surveys, and there are even questions on the necessity of their existence in the future. Our theoretical results presented in this paper call for probability survey samples with rich information on auxiliary variables. A few large scale high quality probability surveys representing the target population can play significant roles in analyzing data from non-probability survey samples. Also,  statistical analysis of  combined or fused data is an important problem \citep{zhang2019}. Further research on this direction will be presented  elsewhere. 

{\red 
\section{Supplementary Material} 

The online supplementary material provides proofs of the Theorems 1 and 2, results from a sensitive analysis, and additional information about the three samples used in the application. 

\section*{Acknowledgments}

The authors are grateful to  the two anonymous referees, the AE, and the Editor for their very constructive comments. The research of J.K. Kim was partially supported by a grant from US National Science Foundation (MMS-1733572). The research of C. Wu was supported by a grant from the Natural Science and Engineering Research Council of Canada. 
}

\newpage 

\begin{table}[H]
\caption{Monte Carlo bias, variance, and relative mean square error (ReMSE)}
\label{table:2}\par
\vskip .2cm
\centerline{\tabcolsep=3truept\begin{tabular}{c|c|r|r|r|r|r|r|r|r|r}
\hline
\multirow{2}{*}{} & \multirow{2}{*}{} & \multicolumn{3}{c|}{Model I} & \multicolumn{3}{c|}{Model II} & \multicolumn{3}{c}{Model III} \\
\cline{3-11}
$n_B$ & Estimator & Bias & Var \ \  & ReMSE & Bias & Var \ \  & ReMSE & Bias & Var \ \  & ReMSE \\
      &           &      & ($\times 10^3$) & ($\times 10^2$)  &  & ($\times 10^3$) & ($\times 10^2$) &  & ($\times 10^3$) & ($\times 10^2$)\\
\hline
 & $\hat{\theta}_A$               & 0.00  & 9.54 & 100     & 0.00 & 9.50 & 100    & 0.00 &  10.81  & 100 \\
  & $\hat{\theta}_{B}$           & -0.64 & 4.68 & 4,299 & -0.32 & 7.98 & 1,151 & -0.64 &  4.43  & 3,791 \\
$500$ & $\hat{\theta}_{I}$              & 0.00  & 10.03 & 105 & 0.00    & 10.26  & 108  & -0.06 & 10.32   & 128 \\
  & $\hat{\theta}_{IPW}$          & -0.03 &  13.81 & 156& -0.01   &  11.27 & 121  & -0.04 & 21.03   & 205 \\
\hline
             & $\hat{\theta}_A$  & 0.00  & 9.69 & 100   & 0.00 & 9.87  & 100  & 0.00 & 10.62   & 100  \\
  & $\hat{\theta}_{B}$      & -0.64 & 2.22 & 4,266 & -0.32 & 4.03  &1,065 & -0.64 & 22.13   & 3,861 \\
$1,000$ & $\hat{\theta}_{I}$            & 0.00  &  8.93 & 93   & 0.00  & 6.20  & 63  & -0.06 & 8.61   & 118 \\
  & $\hat{\theta}_{IPW}$         & -0.04 & 11.77 & 137 & -0.02  & 6.82 & 72    & -0.04 &  16.11  & 167  \\
\hline
\end{tabular}}
\end{table}

\begin{table}[H]
\caption{Monte Carlo mean and relative biase (R.B.) of variance estimators}
\label{table:3}\par
\vskip .2cm
\centerline{\tabcolsep=3truept\begin{tabular}{c|c|c|c|c|c}
\hline
& & \multicolumn{2}{c|}{Linearization} & \multicolumn{2}{c}{Bootstrap} \\
\cline{3-6}
$\;\;$ $n_B$ $\;\;$ & $\;\;$ Model $\;\;$ & $\;\;$ Mean $\;\;$ & $\;\;$ R.B.$\;\;$  & $\;\;$ Mean $\;\;$ & $\;\;$  R.B. $\;\;$ \\
\hline
             &  I  & 0.0102 & 0.019 & 0.0103 & 0.023 \\
$500$       & II  & 0.0109 & 0.067 & 0.0110 & 0.069   \\
& III   & 0.0099 & -0.036 & 0.0106 & 0.026 \\
\hline
& I    & 0.0091 & 0.017 & 0.0091 & 0.019  \\
$1,000$ & II  & 0.0064 & 0.039 & 0.0065 & 0.041  \\
 & III    & 0.0082 & -0.043 & 0.0086 & -0.009  \\
\hline
\end{tabular}}
\end{table}

\newpage

\begin{table}[H]
\caption{Mass imputation estimates of population means by different probability samples}
\label{table:6}\par
\vskip .2cm
\centerline{\tabcolsep=3truept\begin{tabular}{c|c|c|c|c|c|c}
\hline 
$\;\;$ Response $\;\;$ & $\;\;$ PRC $\;\;$ & $\;\;$ CPS $\;\;$ & $\;\;$ BRFSS&BRFSS\textsuperscript{(1)} $\;\;$ & $\;\;$ BRFSS\textsuperscript{(2)} $\;\;$ & $\;\;$ BRFSS\textsuperscript{(3)} $\;\;$ \\\hline 
$Y_1$ &0.461&0.458&0.457& 0.456&0.458 &0.468\\\hline 
$Y_2$ &0.210&0.206&0.200&0.200& 0.203&0.210\\\hline 
$Y_3$ &0.141&0.135&0.133&0.133&0.134&0.137\\\hline 
$Y_4$ &5.301&4.986&4.931&4.921& 4.931&4.997\\\hline 
\end{tabular}}
\end{table}

\begin{table}[H]
\caption{
Variance and variance components of the mass imputation estimators
}
\label{table:7}\par
\vskip .2cm
\centerline{\tabcolsep=3truept\begin{tabular}{ c|c|c|c|c|c|c}
\hline 
 $Y$ &  Method &CPS&BRFSS & BRFSS\textsuperscript{(1)} & BRFSS\textsuperscript{(2)}& BRFSS\textsuperscript{(3)}\\\hline 
 \multirow{3}{*}{$Y_1$}  &  \multirow{2}{*}{L } &$ 4.195$& $4.323$&4.859 &11.848  &75.193 \\
 &&{(0.087, 4.108)}&(0.144, 4.179)&(0.763, 4.096)&(7.732, 4.116) &(70.226, 4.966)\\
  &  Boot  &$4.055$& $4.187$&5.031&12.170&72.559\\\hline 
 \multirow{3}{*}{$Y_2$}  &  \multirow{2}{*}{L} &$2.602$&$2.599$&2.729 &4.737 &23.802 \\
 &&{(0.037, 2.565)}&(0.039,  2.559)& (0.211, 2.518)&(2.121, 2.616) &(20.495, 3.307)\\
 &  Boot &$2.607$&$2.615$&2.822&4.532& 22.840\\\hline 
  \multirow{3}{*}{$Y_3$}  & \multirow{2}{*}{L} &$1.922$&$1.910$&1.950 &2.662 &10.626 \\
  &&{(0.016, 1.906)}&(0.016,  1.894)& (0.083, 1.867)&(0.791, 1.871)& (8.211,  2.415)\\
 &  Boot &$1.930$&$1.886$&1.942& 2.719&10.818\\\hline 
 \multirow{3}{*}{$Y_4$}  &  \multirow{2}{*}{L} &$0.978$&$1.010$& 1.062 &1.785 &8.887\\
 &&(0.012, 0.966)&(0.016, 0.994)&(0.085, 0.977)& (0.818, 0.967)&(7.744, 1.143)\\
 &  Boot &$0.952$ &$0.996$&1.091& 1.808&8.759\\\hline 
\end{tabular}}
\vspace{1ex}
{\raggedright L: Linearization; Boot: Bootstrap. \par}
\end{table}

\begin{table}[H]
\caption{P-values for covariates in Model $\xi ({\rm All})$ }
\label{table:8}\par
\vskip .2cm
\centerline{\tabcolsep=3truept\begin{tabular}{l|llll|llll}
\hline
&\multicolumn{4}{c|}{CPS} &  \multicolumn{4}{c}{BRFSS}\\
 \hline 
 Covariates & $Y_1$& $Y_2$ & $Y_3$& $Y_4$& $Y_1$& $Y_2$ & $Y_3$& $Y_4$ \\\hline
   intercept 	& $*$		&	$*$	&	$*$	&	$*$	&	$*$	&	$*$	&	$*$	&	$*$\\	\hline
 Age 	&	$*$	&	$*$	&	0.188	&	$*$	&	$*$	&	0.044	&	0.305	&	0.071\\	\hline
Female	&	$*$	&	0.990	&	$*$	&	$*$&	$*$	&	0.175	&	$*$	&	$*$	\\	\hline
White only	&	0.010	&	0.163	&	0.665	&	0.066	&	0.064	&	0.074	&	0.879	&	0.086\\	\hline
Black only	&	$*$	&	$*$	&	$*$	&	0.589	&	0.002	&	$*$	&	$*$	&	0.805\\	\hline
Hispanic/Latino	&	0.122	&	$*$	&	$*$	&	0.335&	0.164	&	0.001	&	$*$	&	0.108	\\	\hline
Northeast 	&	$*$	&	0.514	&	0.068	&	0.987	&	$*$	&	0.617	&	0.125	&	0.688\\	\hline
 South	&	0.007	&	$*$	&	0.252	&	0.985	&	0.011	&	$*$	&	0.461	&	0.933\\	\hline
West	&	0.087	&	0.001	&	0.668	&	0.060	&	0.032	&	$*$	&	0.908	&	0.013\\	\hline
Married	&	0.021	&	$*$	&	$*$	&	0.769	&	0.142	&	$*$	&	$*$	&	0.030\\	\hline
 Working	&	0.864	&	0.062	&	0.010	&	$*$	&	0.406	&	0.001	&	$*$	&	$*$\\	\hline
Retired 	&	0.007	&	0.912	&	0.002	&	$*$&	$*$	&	0.006	&	$*$	&	$*$	\\	\hline
High school or less	&	0.364	&	0.134	&	0.304	&	0.016	&	0.314	&	$*$	&	$*$	&	$*$\\	\hline
Bachelor's degree and above	&	0.023	&	0.039	&	0.189	&	0.005&	0.038	&	$*$	&	$*$	&	$*$	\\	\hline
Presence of child in household	&	NA	&	NA	&	NA	&NA&	$*$	&	$*$	&	$*$	&	$*$\\		\hline
Home ownership	&	NA	&	NA &		NA &	 NA	&	0.036	&	$*$	&	$*$	&	0.001	\\\hline
Smoke everyday	&	NA	&	NA	&	NA	&	NA	&	0.847	&	0.893	&	0.067	&	0.311	\\\hline
Smoke never	& NA		&	NA	&	NA	&	NA	&	$*$	&	0.031	&	$*$	&	$*$	\\\hline
No money to see doctors	&	NA	&	NA	&	NA	& NA		&	$*$	&	$*$	&	$*$	&	0.018	\\\hline
Having medical insurance	&	NA	&	NA	&	NA	&	NA	&	0.006	&	$*$	&	0.009	&	0.338	\\\hline
Household income $<$ 20K	&	0.003	&	0.025	&	0.038	&	0.003	&	NA	&	NA	&	NA	&	NA\\	\hline
Household income $>$ 100K	&	0.075	&	0.057	&	0.213	&	$*$&	 NA	&	NA&	NA	&NA	\\	\hline
Volunteered 	&	$*$	&	$*$	&	$*$	&	0.438&NA	&NA	&NA	&NA\\	
\hline
\end{tabular}}
\vspace{1ex}
{\raggedright ``$*$'' indicates that p-value  $< 0.001$. \par}
\end{table}

\begin{table}[H]
\caption{P-values for covariates in Model $\xi({\rm Select})$ }
\label{table:9}\par
\vskip .2cm
\centerline{\tabcolsep=3truept\begin{tabular}{l|llll|llll}
\hline
&\multicolumn{4}{c|}{CPS} &  \multicolumn{4}{c}{BRFSS}\\
 \hline 
 Covariates & $Y_1$& $Y_2$ & $Y_3$& $Y_4$& $Y_1$& $Y_2$ & $Y_3$& $Y_4$ \\\hline
intercept 	&	$*$	&	$*$	&	$*$	&	$*$	&	$*$	&	$*$	&	$*$	&	$*$	\\\hline
 Age 	&	$*$	&	$*$	&	\xmark		&	$*$	&	$*$	&	0.044	&	 \xmark	&	0.072	\\\hline
Female	&	$*$	&	\xmark		&	$*$	&	$*$	&	$*$	&	\xmark		&	$*$	&	$*$	\\\hline
White only	&	0.009	&	\xmark		&	\xmark	&	0.003	&	0.068	&	0.074	&	\xmark	&	0.011	\\\hline
Black only	&	$*$	&	$*$	&	$*$	&	\xmark	&	0.002	&	$*$	&	$*$	&	\xmark	\\\hline
Hispanic/Latino	&	0.118	&	$*$	&	$*$	&	\xmark	&	\xmark	&	0.001	&	$*$	&	0.094	\\\hline
Northeast 	&	$*$	&	\xmark	&	0.013	&	\xmark	&	$*$	&	\xmark	&	0.095	&	\xmark	\\\hline
 South	&	0.007	&	$*$	&	0.095	&	\xmark	&	0.009	&	$*$	&	\xmark	&	\xmark	\\\hline
West	&	0.097	&	$*$	&	\xmark	&	0.009	&	0.016	&	$*$	&	\xmark	&	0.003	\\\hline
Married	&	0.020	&	$*$	&	$*$	&	\xmark	&	0.147	&	$*$	&	$*$	&	0.022	\\\hline
 Working	&	\xmark	&	0.027	&	0.005	&	$*$	&	\xmark	&	0.002	&	$*$	&	$*$	\\\hline
Retired 	&	0.002	&	\xmark	&	$*$	&	$*$	&	$*$	&	0.008	&	$*$	&	$*$	\\\hline
High school or less	&		\xmark&	0.128	&	\xmark	&	0.013	&	\xmark	&	$*$	&	0.001	&	$*$	\\\hline
Bachelor's degree and above	&	0.005	&	0.031	&	0.029	&	0.003	&	0.004	&	$*$	&	$*$	&	$*$	\\\hline
Presence of child in household	&	NA	&	NA	&	NA	&	NA	&	$*$	&	$*$	&	$*$	&	$*$	\\\hline
Home ownership	&	NA	&	NA	&	NA	&	NA	&	0.026	&	$*$	&	$*$	&	0.001	\\\hline
Smoke everyday	&	NA	&		NA&	NA&NA	&	\xmark	&	\xmark	&	0.081	&	\xmark	\\\hline
Smoke never	&		NA&	NA	&NA	&	NA&	$*$	&	$*$	&	$*$	&	$*$	\\\hline
No money to see doctors	&	NA	&	NA&		NA&	NA	&	$*$	&	$*$	&	$*$	&	0.010	\\\hline
Having medical insurance	&	NA	&	NA	&	NA	&NA	&	0.004	&	$*$	&	0.008	&	\xmark	\\\hline
Household income $<$ 20K	&	0.002	&	0.027	&	0.021	&	0.001	&	NA	&	NA	&	NA	&	NA	\\\hline
Household income $>$ 100K	&	0.076	&	0.048	&	\xmark	&	$*$	&	NA	&	NA	&	NA	&	NA	\\\hline
Volunteered 	&	$*$	&	$*$	&	$*$	&	\xmark	&	NA	&	NA	&	NA	&	NA	\\\hline
\end{tabular}}
\vspace{1ex}
{\raggedright ``$*$'' indicates that p-value  $< 0.001$. ``\xmark'' indicates that the covariate is not selected by the backward variable selection algorithm. \par}
\end{table}

\begin{table}[H]
\caption{Mass imputation estimates of population means using different sets of covariates
}
\label{table:10}\par
\vskip .2cm
\centerline{\tabcolsep=3truept\begin{tabular}{c|c|c|c|c}
\hline 
\multirow{2}{*}{ $\;\;$ Response $\;\;$}  &\multirow{2}{*}{$\;\;$ Sample $A$ $\;\;$} & \multicolumn{3}{c}{$\hat{\theta}_{I}$} \\ \cline{3-5}
 &  &  $\;\;\;\;$ $\xi({\rm Partial})$ $\;\;\;\;$ & $\;\;\;\;\;\;$ $\xi({\rm All}) $ $\;\;\;\;\;$ & $\;\;\;\;$ $\xi({\rm Select})$ $\;\;\;\;$ \\\hline 
 \multirow{2}{*}{$Y_1$}  &  CPS & $\;\;$ $0.458 \;\; (4.195)$ $\;\;$ & $\;\;$ $0.404 \;\; (4.600)$ $\;\;$ & $\;\;$ $0.402 \;\; (4.149)$ $\;\;$ \\
 &BRFSS& $0.457 \;\; (4.323)$	&	$0.446 \;\; (4.687)$	&$0.447 \;\; (3.952)$\\\hline
 \multirow{2}{*}{$Y_2$}  &  CPS & $0.206 \;\; (2.602)$ & $0.134\;\; (1.425)$	&	$0.134 \;\; (1.388)$\\
 &BRFSS& $0.200 \;\; (2.599)$	&	$0.198 \;\; (2.789)$	&	$0.198 \;\; (2.786)$\\\hline
 \multirow{2}{*}{$Y_3$}  &  CPS & $0.135 \;\; (1.922)$	&$0.088 \;\; (0.978)$	&	$0.087 \;\; (0.893)$\\
 &BRFSS& $0.133 \;\; (1.910)$	&	$0.121 \;\; (1.842)$	&	$0.120 \;\; (1.701)$\\\hline
 \multirow{2}{*}{$Y_4$}  &  CPS & $4.986 \;\; (0.978)$	&	$5.076 \;\; (1.240)$	&	$5.086 \;\; (0.959)$\\
 &BRFSS& $4.931 \;\; (1.010)$	&	$4.812 \;\; (1.028)$	&	$4.807 \;\; (0.984)$\\\hline
\end{tabular}}
\end{table}

\begin{table}[H]
\caption{Estimates of population means by mass imputation and inverse probability weighting
}
\label{table:11}\par
\vskip .2cm
\centerline{\tabcolsep=3truept\begin{tabular}{c|c|c|c}
\hline 
$\;\;$ Response $\;\;$ & $\;\;$ Sample $A$ $\;\;$ & ${\hat{\theta}_{I}}$ & $\hat{\theta}_{\IPW}$ \\
  \hline
\multirow{2}{*}{$Y_1$} &CPS   & $\;\;\;$ $0.402 \;\; (4.149)$ $\;\;\;$ & $\;\;\;$ $0.396 \;\; (4.170)$  $\;\;\;$ \\
  &  BRFSS& $ 0.447 \;\; (3.952)$ & $0.443 \;\; (4.170)$\\
 \hline
    \multirow{2}{*}{$Y_2$} & CPS  &$0.134 \;\; (1.388)$ & $0.136 \;\; (1.309)$ \\
& BRFSS&  $0.198 \;\; (2.786)$ & $0.193 \;\; (2.686)$  \\
   \hline
  \multirow{2}{*}{ $Y_3$}&CPS   & $0.087 \;\; (0.893)$ & $0.088 \;\; (0.829)$\\
& BRFSS&  $0.120 \;\; (1.701)$ & $ 0.120 \;\; (1.735)$ \\
    \hline
 \multirow{2}{*}{  $Y_4$ }&CPS     & $5.086 \;\; (0.959)$ & $5.059 \;\; (0.989)$ \\
& BRFSS&  $4.807 \;\; (0.984)$ & $4.717 \;\; (0.916)$ \\
     \hline
 \end{tabular}}
\end{table}

\bigskip

\newpage 
\bibliographystyle{dcu}
\bibliography{minp5}

\end{document}